# Raman spectroscopic characterization of crater walls formed upon single-shot high-energy femtosecond laser irradiation of dimethacrylate polymer doped with plasmonic gold nanorods

*István Rigó[1], Judit Kámán[1], Ágnes Nagyné Szokol[1], Roman Holomb[1], Attila Bonyár[2], Melinda Szalóki[3], Alexandra Borók[1,2], Shereen Zangana[2], Péter Rácz[1], Márk Aladi[1], Miklós Ákos Kedves[1], Gábor Galbács[5], László P. Csernai[1,6,7], Tamás S. Biró[1], Norbert Kroó[1,8], Miklós Veres[1], NAPLIFE Collaboration*

[1]Wigner Research Centre for Physics, Budapest, Hungary

[2]Department of Electronics Technology, Faculty of Electrical Engineering and Informatics, Budapest University for Economics and Informatics, Budapest, H1111, Hungary

[3]Department of Biomaterials and Prosthetic Dentistry, Faculty of Dentistry, University of Debrecen, Debrecen, Hungary

[4]Centre for Energy Research, Institute of Technical Physics and Materials Science (MFA), H1121 Budapest, Hungary

[5]Department of Inorganic and Analytical Chemistry, University of Szeged, Szeged, H-6720 Hungary

[6]Department of Physics and Technology, University of Bergen, 5007 Bergen, Norway

[7]Frankfurt Institute for Advanced Studies, Frankfurt/Main, Germany

[8]Hungarian Academy of Sciences, 1051 Budapest, Hungary

**Abstract**

The bonding configuration of the crater walls formed in urethane dimethacrylate-based polymer doped with plasmonic gold nanorods upon irradiation with a single-shot high-energy femtosecond laser pulse has been studied by Raman spectroscopy. New Raman bands were detected in the 2000-2500 cm$^{-1}$ region of the Raman spectrum the intensities of which showed strong dependence on the concentration of the plasmonic nanoparticles and the energy of the laser pulse. Based on model calculations of the Raman frequencies of the polymer these peaks were attributed to carbon-deuterium and nitrogen-deuterium vibrations. Their appearance might indicate the occurrence of nuclear reactions in the polymer excited by the ultra-strong laser field amplified by the plasmonic nanoparticles.

*Keywords: plasmonic enhancement, femtosecond laser, nuclear reactions, Raman spectroscopy, dimethacrylate polymer*

1. **Introduction**

The use of high-energy short laser pulses is gaining significance in numerous applications. Localized surface plasmon polaritons (LSPP) can, for example, be excited efficiently with these pulses even up to very high laser intensities. One of the reasons to do so is that with the help of plasmonic nanoparticles the electromagnetic field of the lasers can be amplified on the nanoscale resulting in fields orders of magnitude higher, than those of the laser pulse. Our motivation to explore this enhancement effect has been to use these high fields to realize tabletop plasmonic nano-fusion processes [1]. Ti:Sa femtosecond laser pulses have been used to excite LSPPs in a polymer sheet, containing resonant plasmonic gold nanorods, with laser intensities up to a few times $10^{17}$ W/cm$^2$. The first step in our studies has been the analysis of the craters formed by individual laser shots in the material [2]. It was found, that the volume of these craters is always significantly higher in the polymer containing the gold nanoparticles than in the same polymer without them. The volume increase can be attributed only to nuclear processes, and it has been found, that the crater volumes depend linearly on the energy of the laser pulses [2]. Since this energy production has been found to be comparable, or even larger, than that of the energy of the laser pulse, it has been decided to use vibrational spectroscopic method (Raman spectroscopy) to characterize structural changes occurred in the polymeric structure at the crater walls formed upon irradiation of the material with single-shot high-energy femtosecond laser pulse. The irradiation with different pulse energies was performed on both a pure structure and also samples doped with plasmonic gold nanorods (in two different concentrations) having plasmon resonance at the wavelength of the femtosecond laser, and the Raman measurements were performed on both types of samples.

Being sensitive to bonding configuration, Raman spectroscopy is widely used to characterize different polymeric materials, including dimethacrylates. This method can be used to determine the degree of conversion in these types of polymers, as well, through the changes and ratio of the Raman peak of the C=C double bonds to that of a reference band belonging to a bond not affected by the polymerization [3-5]. In addition to the bonds of the polymer frame it allows to study the different C-H and N-H groups as well. The effect of the laser pulse on the polymer framework, including the additional conversion upon irradiation in the presence of gold nanoparticles, has been published earlier [6]. Here we report on the changes observable in the 2000-2600 cm$^{-1}$ region, being dependent on the presence of the nanoparticles and the pulse energy as well. In situ laser induced breakdown spectroscopy (LIBS) measurements were also performed to validate the Raman findings.

## 2. Methods

*2.1. Materials and sample preparation*

The photopolymerizable dimethacrylate resin mixture consists of urethane dimethacrylate (UDMA) (Sigma Aldrich) and triethylene glycol dimethacrylate (TEGDMA) (Sigma Aldrich) in 3:1 mass ratio. Dodecanethiol-capped gold nanorods (Au-DDT) in size of 25 nm diameter and 75 nm length were purchased from Nanopartz Inc. (part no.: B12-25-700-1DDT-TOL-50-0.25, and B12-25-750-1DDT-TOL-50-0.25).

The preparation method of pure (UDMA-X) and doped with gold nanorods (UDMA-Au) polymer samples is described in detail elsewhere [6]. In short, the mixture of the UDMA and TEGMA monomers and the photoinitiator (and Au-DDT) is placed on a glass slide in a template allowing to obtain a thin layer of the resin. Then the polymerization mixture is irradiated with a standard dental curing lamp emitting blue light with 3 min exposure time. The obtained round shaped thin film samples are clear (UDMA-X non-doped polymer) or have pink color (UDMA-Au doped samples). The gold nanorods were added to the monomer mixture in 2 different concentrations of 0.124 m/m% and 0.182 m/m%, and the corresponding samples were marked as UDMA-Au1 and UDMA-Au2 in this manuscript, respectively. In addition, on some figures, the laser pulse energy in milliJoules is also added to the sample name, i.e. UDMA-X-25 corresponds to the crater in the non-doped polymer irradiated with 25 mJ laser pulse.

## 2.2. Laser irradiation experiments

The irradiation of the samples was implemented by a Ti:Sapphire-based chirped-pulse two-stage amplifier-laser system (Coherent Hydra) delivering pulses with 40 fs pulse length at 795 nm central wavelength with 10 Hz repetition rate and 25 mJ maximum pulse energy. The beam was focused with a lens having 50 cm focal length. The laser irradiation experiments were performed under vacuum conditions to avoid nonlinear processes in air. The pressure in the vacuum chamber was in the range of $10^{-6}$ mbar. The sample treatment was achieved by single pulses of different energy, each illuminating a different region of the sample surface (this was achieved by shifting the sample laterally after each pulse). Since the energy density is above the ablation threshold of the material, crater formation was observed at the irradiation spot.

## 2.3. Raman spectroscopy

A Renishaw InVia micro-Raman spectrometer connected to a LeicaDM2700 microscope was used for the Raman spectroscopic measurements. The Raman spectra were recorded on the wall of the formed crater with 532 nm excitation in backscattering geometry, with the laser focused into a spot of ~2 microns diameter on the sample surface by using a 50X/0.5 NA objective. The laser power was ~6 mW which equals 5-10% of the maximum intensity of the laser source. The spectra were recorded in the 1000 - 2650 cm$^{-1}$ spectral region. Due to the low intensity of the investigated Raman bands the accumulation time was set to 2 hours in each spot. Before the measurements, calibration was done by using a silicon wafer and its characteristic peak at 520 cm$^{-1}$ Raman shift.

Since this study is about the evaluation and comparison of small-intensity Raman peaks recorded on different craters formed upon single-shot laser irradiation, a special care was given to the processing of the measured Raman spectra, that was done with the Origin 2019 software and a custom Matlab code. The first step was the normalization of the spectra to the C-C peak at 1460 cm$^{-1}$. Then, background subtraction was performed by using the Asymmetric Least Squares method [7]. The integral intensity of the Raman peaks was determined by calculating the area under the curve for the composite Raman band observed in the 2000-2600 cm$^{-1}$ region.

*2.4. Laser induced breakdown spectroscopy*

The LIBS measurements were performed in situ, during the single-shot irradiation of the polymer targets with the high-intensity laser pulses. The laser induced breakdown plasma emission was collected at 45 degrees angle and collimated by placing a lens at a 4 cm distance from the target inside the chamber. The collimated beam was then focused into an optical fiber with a core diameter of 400 μm and transmitted into a high resolution double Echelle spectrometer (having a 4 pm resolution) equipped with an ICCD camera (Demon, LTB, Berlin). The spectrum recording was done using a fixed gate width of 1 μs and a gate delay of 0.6 μs. The spectrometer was centered around 656 nm and recorded the spectrum in a 3 nm spectral window.

*2.5. Modeling and calculations of the selectively deuterized UDMA monomer*

The chemical structure of urethane-dimethacrylate (UDMA) monomer is shown in Figure 1. Part of this structure containing N-H and C-H$_2$ groups was selected as a starting geometry for further modeling and subsequent calculations. For better description of the chemical environment, the dangling chemical bonds at the ends of the model were terminated with methyl (CH$_3$) and hydroxyl (OH) groups, as shown in Figure 2.

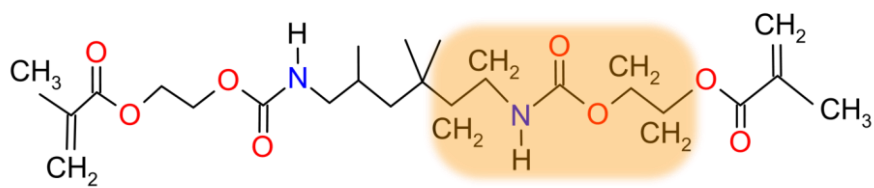

*Figure 1. Chemical structure of UDMA monomer together with the part selected for further modeling and calculations.*

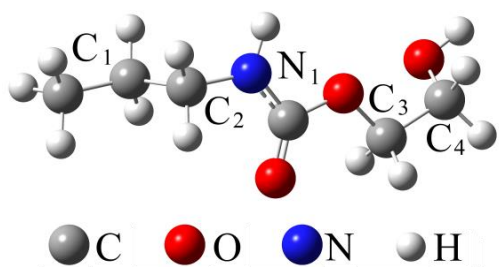

*Figure 2. Optimized (B3LYP/6-311++G(d,p)) geometry of UDMA model ($C_1H_2$-$C_2H_2$ and $C_3H_2$-$C_4H_2$ groups are in anti and gauche conformational states, respectively).*

The self-consistent density functional theory (DFT) field method using the hybrid B3LYP functional consisting of a linear combination of the pure corrected exchange functional by Becke [8] and the three-parameter gradient-corrected correlation functional by Lee *et al.* [9] was applied for geometry optimization of UDMA model and Raman spectra calculations using the Gaussian-09 program package [10]. The triple zeta valence (TZV) Pople 6-311++G(d,p) basis set was used for all atoms [11]. The total energy of the gas-phase model was optimized using the Berny optimization algorithm. The vibrational frequencies of the model calculated using the same method and basis set verified the optimized geometry as true energy minimum structure. The Raman activities of the vibrational modes of the UDMA model were also calculated. To study the effect of selective deuterization on the Raman spectra of the UDMA model, the $C_1H_2$, $C_2H_2$, $C_3H_2$, $C_4H_2$ and $N_1H$ groups (hereafter denoted as $C_1$, $C_2$, $C_3$, $C_4$ and $N_1$ sites) were used for hydrogen-to-deuterium substitution (Figure 2). The optimized structure of the UDMA models was used for further vibrational mode frequency calculations of selectively deuterated structures. For this reason the exact mass of the deuterium isotope (2.01410 a.m.u.) was used for hydrogen atoms bonded to the selected site (center).

To simulate the Raman spectra of the models a Lorentz-shape function with the intensity proportional to the calculated Raman activity and with full width at half-height (FWHH) of 20 cm$^{-1}$, modeling the natural bandwidth of the experimental spectra, was applied for each of the computed Raman modes.

3. **Results and discussion**

*3.1. Differences in the Raman spectrum of the crater wall in samples with and without gold nanorods*

Figure 3 compares the normalized Raman spectra recorded in a non-irradiated spot and on the crater wall of doped (UDMA-Au1) and non-doped (UDMA-X) samples. Between 1400 and 2000

cm$^{-1}$ all the spectra are dominated by the characteristic Raman peaks of the UDMA:TEGDMA polymer [6]. A closer look shows, however, that in the spectrum of the crater walls the bands at 1640 and 1715 cm$^{-1}$ (and also at 1400 cm$^{-1}$) decrease remarkably, and this effect is more pronounced for the higher pulse energies. These peaks are related to C=C and C=O vibrations of the methacrylate group and their change indicates additional polymerization of the sample caused by the laser irradiation. This behavior has been studied by us earlier and the results are published elsewhere [6].

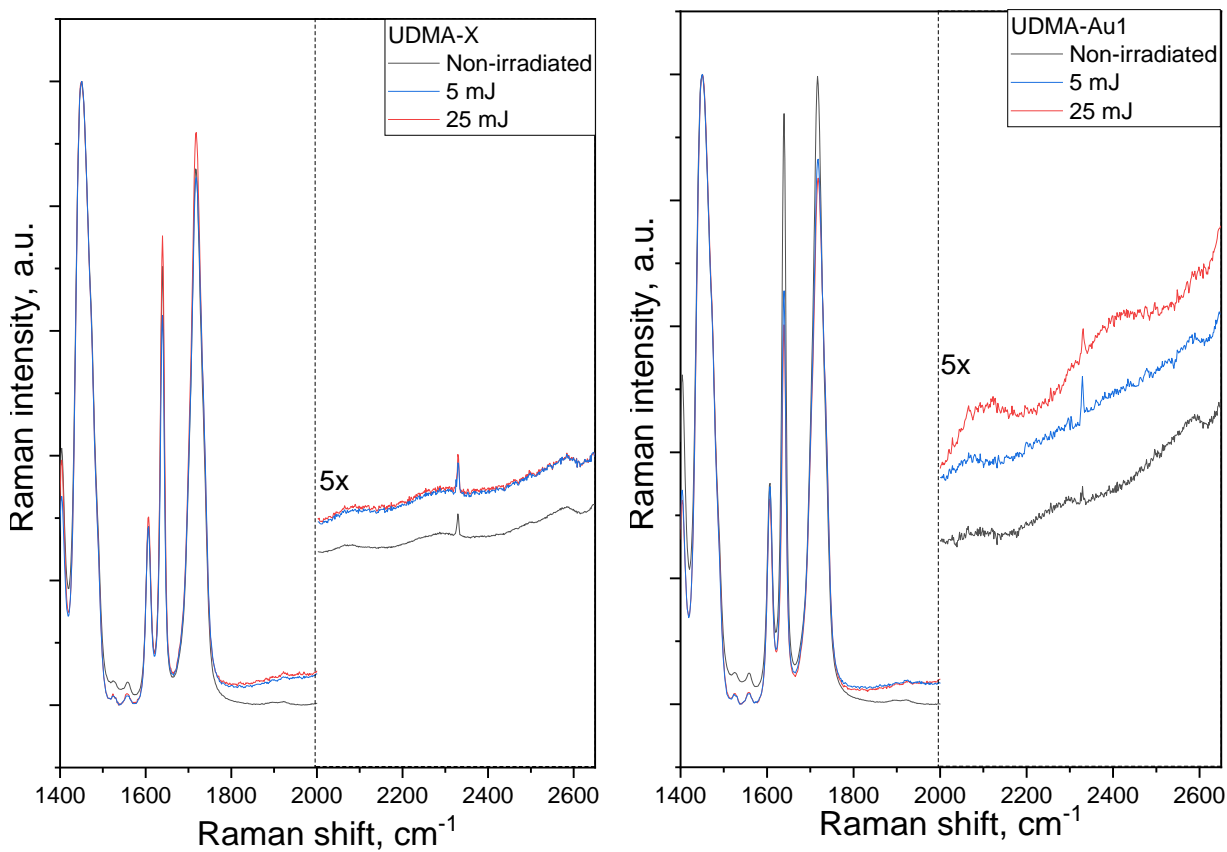

*Figure 3. Raman spectra recorded in a non-irradiated spot and on crater walls formed upon single-shot pulsed laser irradiation of UDMA polymer samples without (UDMA-X - left) and with (UDMA-Au1 - right) gold nanorods with ultrashort laser pulses of different energy. The left part of each graph shows the 2000-2650 cm$^{-1}$ region enlarged for clarity.*

In addition to the above, some changes can be observed in the 2000-2650 cm$^{-1}$ region. Here some weak and broad Raman bands can be observed in the spectrum of the non-irradiated polymer at 2075, 2295, 2500 and 2580 cm$^{-1}$ (and a narrow peak at 2330 cm$^{-1}$ arising from the nitrogen molecules being present in the air above the sample surface). After the irradiation the spectra of both the undoped and doped samples indicate an increased PL background, the level of which is dependent on the pulse energy for the UDMA-Au1 sample. The intensity and the shape of the weak Raman bands recorded on the non-irradiated spot and the crater wall of the UDMA-X sample are very similar. In contrast, a well-noticable extra Raman contribution can be observed in the spectra recorded on the crater walls of the doped UDMA-Au1 sample. Figure 4 compares the differential Raman spectra obtained by subtracting the spectrum of the non-irradiated polymer from those recorded on the crater walls formed upon single-shot irradiation of the UDMA polymer with and without gold nanorods. The Figure focuses on the 1950-2650 $cm^{-1}$ spectral region. Both the UDMA-X and UDMA-Au spectra show the additional PL background, which is very similar to the UDMA-X sample irradiated with the two different pulses, but it is some 6 and 10 times higher in the spectra of the doped sample and 5 mJ and 25 mJ pulse energies, respectively. Some well-resolved bands can also be observed in the differential spectra of the UDMA-Au sample, indicating additional Raman contribution by species formed during the laser-matter interaction. Two broad peaks can be seen with maxima around 2120 $cm^{-1}$ and 2400 $cm^{-1}$, and their intensity depends on the laser pulse energy. These features are difficult to observe in the spectra of the single-shot irradiated non-doped samples.

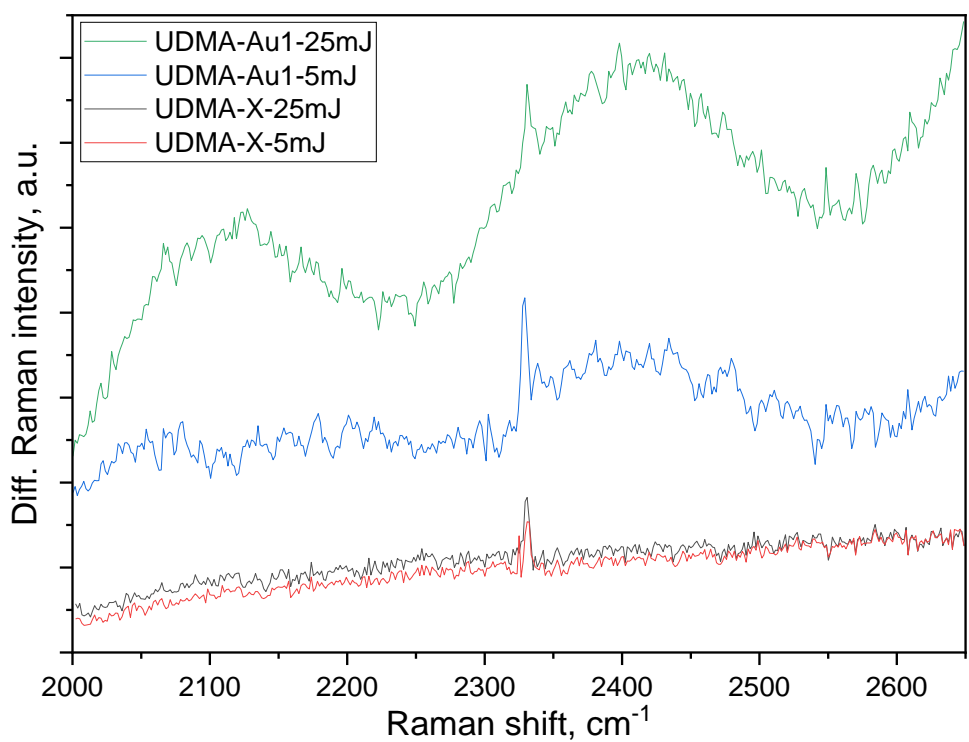

*Figure 4. Differential Raman spectra of the crater walls and the non-irradiated polymer structure for 5 and 25 mJ pulse energies.*

Since the two UDMA samples used for the irradiation experiments were fabricated from the same monomer mixture, and the only difference is the addition of the gold nanorods, it can be assumed that the nanoparticles are involved in the appearance of the two Raman peaks in the spectra of the UDMA-Au sample.

Another set of experiments was performed with samples doped with gold nanorods in two different concentrations and a non-doped UDMA-X reference sample. The irradiation was performed with 4 different pulse energies ranging from 2.5 to 25 mJ. The dependence of the integrated intensity (after background subtraction) in the 2000-2580 cm$^{-1}$ spectral region of the two Raman bands described above on the laser pulse energy is shown in Figure 5. Each value in the Figure corresponds to the average of five Raman spectra recorded in the same crater.

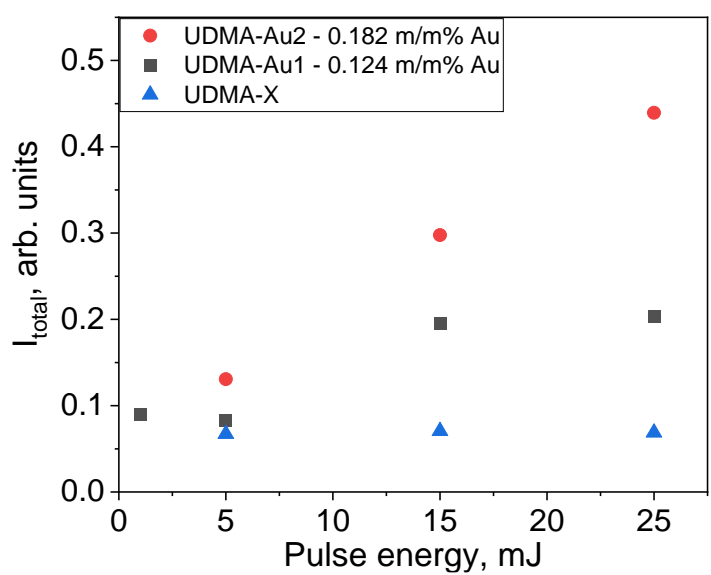

*Figure 5. Dependence of the integrated intensity of the broad Raman band in the 2000-2580 cm$^{-1}$ region of the Raman spectra of the crater walls on the laser pulse energy. Each point represents an average obtained on five Raman spectra recorded in the same crater.*

It can be seen that for the 5 mJ UDMA-X sample the peak intensity is 0.067, and it does not change much with the pulse energy (0.069 for the 25 mJ crater). Some variation can be observed for the UDMA-Au1 sample, where the band intensity slightly decreases up to 5 mJ (to 0.083, yet above the value for the UDMA-X sample), but then increases up to 0.204 for the 25 mJ crater, showing a clear tendency. The increase of the peak area with the pulse energy is more pronounced for the UDMA-Au2 sample, where it starts from 0.131 for 5 mJ and reaches 0.459 at 25 mJ. The observed behavior also supports that the new peaks appearing in the 2000-2500 cm$^{-1}$ region are related to the presence and the concentration of the gold nanorods.

The experiments were repeated with a third set of doped and undoped UDMA samples, with 25 mJ irradiation of the three polymers, in 6 different spots each. The integrated peak intensities of the Raman spectra for the 2000-2580 cm$^{-1}$ region are compared in Figure 6. It can be seen that the six craters formed in the UDMA-X sample show the lowest Raman band intensities with minimal fluctuations among the craters (0.065 ±0.001). Both UDMA-Au1 and UDMA-Au2 show much higher peak intensities and larger crater-to-crater variance (0.210 ±0.018 and 0.443 ±0.034, respectively). These findings are in good agreement with the earlier observations (Fig. 5).

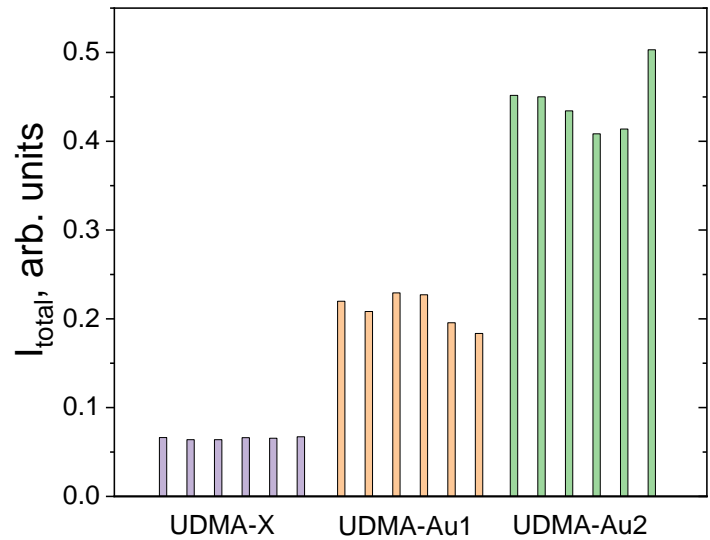

*Figure 6. The integrated intensity of the broad Raman band in the 2000-2500 cm$^{-1}$ region of the Raman spectra, recorded on separate craters formed upon irradiation of the samples with 25 mJ laser pulse energy.*

The small variance in the Raman peak intensities in the undoped UDMA-X craters indicates the high homogeneity of the polymer structure and uniformity of the single-shot laser irradiation. However, the doped polymers contain the gold nanoparticles in a relatively small density (60-120 μm$^{-3}$), and the Raman spectra with 1.3 micron spot size (corresponding to 0.55 μm$^3$) could be recorded in areas with different gold nanoparticle amounts. As a consequence, the band intensities will fluctuate in a broader range.

Based on the analysis of the Raman spectra it can be assumed that the appearance of the broad Raman band upon single-shot femtosecond laser irradiation is related to the presence of gold nanoparticles. They could play different roles in this phenomenon: i) increase the Raman sensitivity due to the surface-enhanced Raman scattering (SERS); (ii) interact with the species being present in the UDMA structure and; (iii) facilitate structural transformations in the polymer through the plasmonic effect.

The SERS effect in our measurements is of very low probability. We used 532 nm excitation, and with this, the Raman bands of UDMA can be detected in the 540-640 nm region (500-3200 cm$^{-1}$). This spectral region is, however, far from the 795 nm resonant wavelength of the gold nanoparticles.

Gold shows low activity in forming chemical bonds with carbon, nitrogen, oxygen and hydrogen atoms, being present in the polymer. But it should be noted that the existence of Au-alkyl bond has been experimentally evidenced recently by using a SERS active surface. However, the corresponding Raman mode at ~387 cm$^{-1}$ characteristics of Au-C bond stretching vibrations is located in the low-frequency spectral region [12]. Taking into account this fact and the masses of N and O atoms, the hypothetical Au-N and Au-O bonds of possible radiation-induced complexes can be excluded from the analysis. Therefore, the only Au-H stretches among hypothetic Au-X (X=C, N, O and H) bonds can potentially contribute to the spectral region of our interest (2000-2500 cm$^{-1}$). Such surface Au-H species were found to be generated upon the chemisorption of molecular hydrogen on supported Au catalysts. For instance, the IR spectra of gold nanoparticles supported on nanoparticulate ceria (Au/CeO$_2$) show a new band around 2130 cm$^{-1}$ assigned to Au-H stretching vibration [13, 14]. Other absorption measurements performed on laser-ablated gold atoms after the reaction with dihydrogen report the Au-H band at 2164 cm$^{-1}$ [15]. As a consequence, the formation of Au-H species cannot be completely excluded.

The plasmonic effect of the nanorods on structural transformations, on the other hand, is of high probability. Since the plasmon resonance of the nanoparticles is tuned to the wavelength of the laser used for the single-shot treatment, these nanostructures act as nanoantennas and amplify the electromagnetic field of the incident light in their surroundings. This enhancement could reach a few orders of magnitude, and presumably, this field, being much stronger here than in the undoped UDMA, is responsible for the difference in the structural transformations and the appearance of the new Raman features in the 2000-2500 cm$^{-1}$ region.

On the polymer side, the additional Raman scattering contribution appearing in the 2000-2500 cm$^{-1}$ region of the Raman spectrum of the craters formed upon irradiation of the UDMA samples could have three possible origins: it could 1) arise from some contamination, 2) be not a Raman but a photoluminescence feature or 3) belong to some new structural units. The contamination of the polymer can be excluded, since the UDMA-X, UDMA-Au1 and UDMA-Au2 samples were prepared from the same monomer mixture in each of the two sample sets, but the peaks are of much lower intensity in the UDMA-X sample. The gold nanorods, being another possible contamination source, were of high purity and two different batches were used for the UDMA-Au1 and UDMA-Au2 sample sets. In addition, in case of their involvement the additional Raman signal would appear in the spectrum of the non-irradiated samples as well.

The widths of the two-component peaks of the broad Raman feature (located around 2100 cm$^{-1}$ and 2400 cm$^{-1}$) are 100-150 cm$^{-1}$ which corresponds to 3.7-5.5 nm on the wavelength scale.

Photoluminescence peaks of this small width are characteristic for molecules and not for bulk polymeric samples, so this origin can be excluded as well.

The third possible assignment of the new peaks is the formation of new structural units. The host UDMA polymer contains different C-H, N-H, C=C, C-O, and C=O bonds and corresponding functional groups. None of these could give Raman peaks in the 2000-2500 cm$^{-1}$ region. With the elements present in UDMA, the alkyne (C≡C), nitrile (C≡N), azide (N=N=N), and carbon-deuterium(C-D) bonds only were found to have Raman bands in the 2000-2300 cm$^{-1}$ region [15] (and the (H$_2$)AuH group discussed above). The alkyne bond is the least stable of the possible C-C bonds, therefore, its formation upon laser irradiation is of small probability. The same is true for the azide group: since the UDMA monomer contains only two nitrogen atoms per monomer, the conditions are not favorable for the formation of triple nitrogen chains. Formation of nitrile is possible, but this bond has a Raman peak around 2220-2260 cm$^{-1}$, which is right in-between the band positions observed in the Raman spectra (see Fig. 3). The different C-D bonds, however, have their Raman peak in the 2000-2200 cm$^{-1}$ region, which is in good agreement with the position of one of the peaks observed in the Raman spectrum. This assignment suggests that deuterium atoms were formed during the interaction of the ultrashort laser pulse with the polymeric structure, and this effect is more pronounced when the local electromagnetic field is enhanced by the plasmonic gold nanorod antennas. As a consequence, carbon-deuterium groups form, and the corresponding peak can be detected in the Raman spectrum in the 2000-2200 cm$^{-1}$ region. Assuming this origin, the deuterons will be attached not only to the carbon atoms in the UDMA frame, but also to nitrogen. A simple estimation of the shift of the N-H Raman peak position at 3465 cm$^{-1}$ upon the N-H to N-D substitution gives the latter to be around 2500 cm$^{-1}$, which is in good agreement with the position of the second peak observed in the Raman spectrum (see Fig. 4). In order to verify this hypothesis, density functional theory calculations were performed to determine the Raman peak positions of the different C-D and N-D structural units.

### 3.2. LIBS measurements

Typical LIBS spectra recorded on the UDMA-X and UDMA-Au2 samples are shown in Figure 7. Both the H-α line at 656.29 nm and the D-α line at 656.11 nm can be observed in the case of the polymer doped with Au nanoparticles, but for the pristine polymer the D line is absent. In addition, several other emission bands can be observed in the spectra, that could arise from trace contamination in the samples. The LIBS measurements show a clear correlation between the D-α line and the presence of Au NPs is, and support the deuterium formation during the high-intensity laser irradiation in the presence of the plasmonic nanoparticles. A more detailed report on the LIBS measurements performed on these is provided in [16].

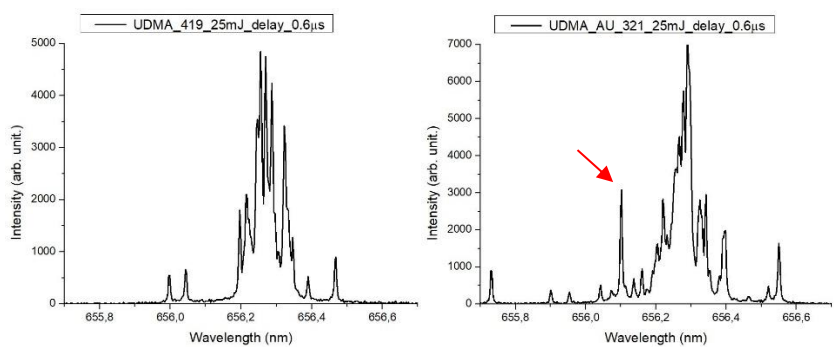

*Figure 7. Typical LIBS spectra of the UDMA-X and UDMA-Au2 samples in the region of the H-α and D-α emission peaks. The red arrow marks the emission peak of deuterium.*

*3.3. DFT calculations of UDMA monomer during H-to-D substitution*

Figure 8 shows the spectral region characteristic of C-H and N-H stretching vibrations in the simulated Raman spectra of the UDMA model (see Fig. 2) in their natural and different selectively deuterated states. The C-H vibrations are in the 2800-3000 cm$^{-1}$ region, while the N-H band is around 3465 cm$^{-1}$. The tentative band assignments are given in Table 1 and Figure 9. It can be seen that the bands in the 2800-3000 cm$^{-1}$ region are related to symmetric and asymmetric vibrations of the $CH_2$ group being in different bonding configurations. The NH group has only one vibrational frequency.

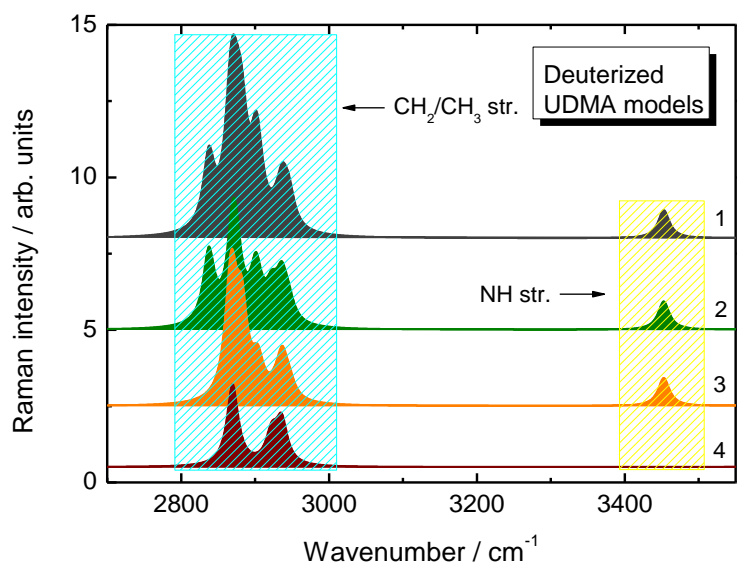

*Figure 8. The spectral region of CH and NH stretching vibrations in the simulated Raman spectra of the UDMA model (Fig. 2) in natural state (1) and selectively deuterated states at $C_1$-$C_2$ (2), $C_3$-$C_4$ (3) and $C_1$-$C_2$, $C_3$-$C_4$ and $N_1$ sites (4). The simulated spectra were frequency corrected by using the scaling factor of 0.95.*

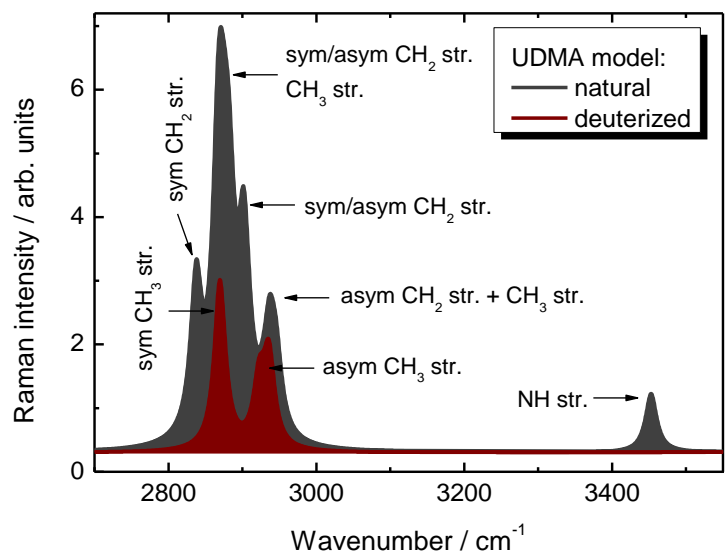

*Figure 9. Assignment of the Raman peaks obtained in the simulated Raman spectra of natural and fully deuterated ($C_1$-$C_2$, $C_3$-$C_4$ and $N_1$ sites) UDMA models. The simulated spectra were frequency corrected by using the scaling factor of 0.95.*

*Table 1. Calculated (6-311++G(d,p)) frequencies (v, cm$^{-1}$) and Raman activities (RA, Å$^4$/a.m.u.) of vibrational modes of natural (H) and fully deuterized (D) UDMA models together with mode assignments. All the calculated mode frequencies were corrected by using a scaling factor of 0.95.*

| $v_D$ | $RA_D$ | $v_H$ | $RA_H$ | Assignment |
|---|---|---|---|---|
| 2062 | 69.4 | 2837 | 147.1 | *sym* C(H/D)$_2$*str.* (OH) |
| 2088 | 59.6 | 2867 | 165.9 | *sym* C(H/D)$_2$*str.* (CH$_3$) |
| 2101 | 48.6 | 2882 | 169.2 | *sym* C(H/D)$_2$*str.* (NH) |
| 2108 | 74.3 | 2901 | 118.0 | *sym* C(H/D)$_2$*str.* (OC) |
| 2136 | 45.6 | 2875 | 96.7 | *asym* C(H/D)$_2$*str.* (OH) |
| 2155 | 42.1 | 2904 | 75.4 | *asym* C(H/D)$_2$*str.* (CH$_3$) |
| 2184 | 10.3 | 2946 | 31.1 | *asym* C(H/D)$_2$*str.* (NH) |
| 2188 | 21.2 | 2945 | 35.1 | *asym* C(H/D)$_2$*str.* (OC) |
| 2530 | 34.0 | 3453 | 68.4 | N(H/D) *str.* |
| 2530* | 35.1 | 3453 | 68.4 | $^{14}$N(H/D) *str.* |
| 2518* | 35.1 | 3445 | 68.4 | $^{15}$N(H/D) *str.* |

*sym str.* and *asym str.* - symmetric and asymmetric stretching vibrations, respectively;

* vibrational mode frequencies calculated natural and deuterized at N-site UDMA model using $^{14}$N and $^{15}$N isotopes.

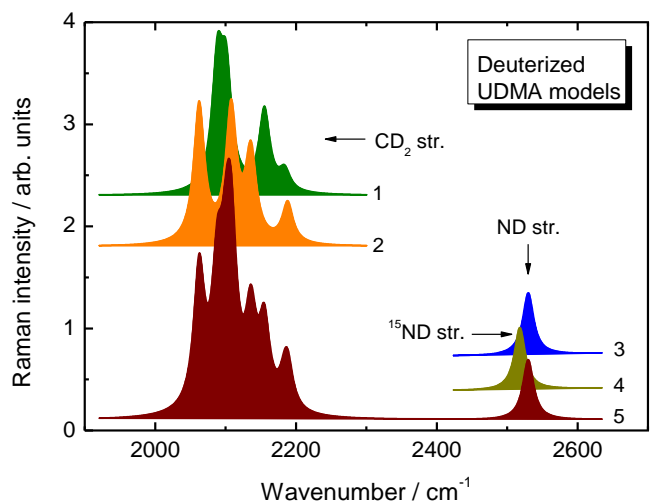

Figure 10. Spectral region of CD and ND stretching vibrations in the simulated Raman spectra of the UDMA model used for selective H-to-D substitution at $C_1D_2$-$C_2D_2$ (1), $C_3D_2$-$C_4D_2$ (2), $N_1D$ (3) and $^{15}N_1D$ (4) groups together with the fully (all groups) deuterized state (5). The simulated spectra were frequency corrected by using the scaling factor of 0.95.

Figure 10 shows the simulated Raman spectra of UDMA models selectively deuterated at $C_1$-$C_2$, $C_3$-$C_4$ and $N_1$ sites (see Fig. 2) in the region of C-D and N-D stretching vibrations. As can be seen, the selective H-to-D substitution at C sites of UDMA models leads to the appearance of bands in the region of 2000-2200 cm$^{-1}$, characteristics of $CD_2$ stretching vibrations. In addition to the location of a particular $CD_2$ group (*i.e.* nearest neighbor atoms), the frequency positions of these bands are found to be dependent on the conformational state (gauche and anti) of $CD_2$ groups (Figure 10, curves 1 and 2). The Raman band at 2530 cm$^{-1}$ characteristics of N-D stretching vibrations was calculated for UDMA model deuterized at $N_1$ site (Figure 10, curve 3). The superposition of C-D stretching vibrations of different $CD_2$ groups form the broadband in the 2000-2200 cm$^{-1}$ region of simulated Raman spectra of fully ($C_1$-$C_2$, $C_3$-$C_4$ and $N_1$ sites) deuterized UDMA model are shown in curve 5 of Figure 10. The assignment of the Raman bands is given in Table 1. It should be noted that in case of the natural model some of the CH modes are coupled, and for them the frequency change is not monotonic/proportional during H-to-D substitution.

The main reason for the broadening of the Raman band, characteristic of the ND stretching vibrations in the Raman spectra of irradiated UDMA sample is due to the disordering effect, accompanied by different types of third-order interactions of ND groups with the surrounding atoms. Also, the natural nitrogen consists of two stable isotopes: the vast majority (~99.6%) of naturally occurring nitrogen is nitrogen-14. The average abundance of $^{15}N$ in air is a very constant 0.366% [17]. In case of $^{15}N$ isotope (exact mass is 15.00011 a.m.u.), the Raman band at 3453 cm$^{-1}$ characteristics of $^{14}$N-H stretching vibrations of UDMA model is red-shifted to 3445 (8 cm$^{-1}$)

(Table 1). The slightly bigger shift (12 cm$^{-1}$) due to N isotope effect is observed for N-D stretching vibrations of UDMA model deuterized at $N_1$ site, where the Raman bands at 2530 and 2518 cm$^{-1}$ were calculated for $^{14}$N and $^{15}$N isotopes, respectively (Table 1., Figure 10, curve 4). Therefore, taking into account that a small amount of $^{15}$N isotope can be present in the structure of UDMA monomer, the N-isotope effect can be an additional reason for the broadening of the ND stretching vibrational mode in the Raman spectra of the irradiated sample.

**Conclusions**

Raman spectroscopic measurements were performed on the crater walls formed in urethane dimethacrylate - triethylene glycol dimethacrylate polymer containing plasmonic gold nanorods upon irradiation with a single-shot femtosecond laser pulses of different energy. New Raman bands were detected in the 2000-2500 cm$^{-1}$ region of the Raman spectrum the intensity of which shows strong dependence on the presence/concentration of the plasmonic nanoparticles and the energy of the laser pulse. Based on model calculations of the Raman frequencies of the polymer these peaks were attributed to carbon-deuterium and nitrogen-deuterium vibrations and their appearance indicates the occurrence of nuclear processes involving the polymer in the laser field amplified by plasmonic nanoparticles, forming hotspots at their surface. The presence of deuterium in the plasma formed during the laser irradiation of the polymer containing plasmonic nanoparticles was also revealed by laser induced breakdown spectroscopy. These results obtained at these relatively low laser pulse energies (intensities) are encouraging for our plans to continue our studies at much higher energies to reach economically feasible tabletop nuclear fusion.

**Acknowledgments**

This work was supported by Nanoplasmonic Laser Fusion Research Laboratory project financed by the National Research and Innovation Office (NKFIH- 2022-2.1.1-NL-2022-00002) and by the Eötvös Lóránd Research Network (ELKH), Hungary. The research reported in this paper and carried out at the Budapest University of Technology and Economics has been supported by the NRDI Fund (TKP2020 IES, Grant No. BME-IE-BIO) based on the charter of bolster issued by the NRDI Office under the auspices of the Ministry for Innovation and Technology. This work was supported by the VEKOP-2.3.2-16-2016-00011 grant, which is co-financed by the European Union and European Social Fund.